\documentclass[preprint,nofootinbib]{revtex4-1}
\usepackage{amsmath}
\usepackage{graphicx}

\newcommand{\be}{\begin{equation}}
\newcommand{\ee}{\end{equation}}

\newcommand{\bi}{\begin{itemize}}
\newcommand{\ei}{\end{itemize}}

\newcommand{\bea}{\begin{eqnarray*}}
\newcommand{\eea}{\end{eqnarray*}}
\newcommand{\tr}{\hbox{tr}}

\DeclareMathOperator{\sgn}{sgn}

\begin{document}

\title{Topological Index Theorem on the Lattice through the Spectral
  Flow of Staggered Fermions}

\author{V. Azcoiti}
\email{azcoiti@azcoiti.unizar.es}

\author{E. Follana}
\email{efollana@unizar.es}

\author{A. Vaquero}
\email{Alejandro.Vaquero@mib.infn.it}
\altaffiliation{present address: INFN, Sezione di Milano-Bicocca.}
\affiliation{Universidad de Zaragoza.}

\author{G. Di Carlo}
\email{giuseppe.dicarlo@lngs.infn.it}
\affiliation{INFN, Laboratori Nazionali del Gran Sasso.}

\date{\today}

\begin{abstract}
We investigate numerically the spectral flow introduced by Adams for
the staggered Dirac operator on realistic (quenched) gauge
configurations. We obtain clear numerical evidence that the definition
works as expected: there is a clear separation between crossings near
and far away from the origin, and the topological charge defined
through the crossings near the origin agrees, for most configurations,
with the one defined through the near-zero modes of large
taste-singlet chirality of the staggered Dirac operator. The crossings
are much closer to the origin if we improve the Dirac operator used in
the definition, and they move towards the origin as we decrease the
lattice spacing.

\end{abstract}

\pacs{12.38.Gc,11.15.Ha}

\maketitle

\section{Introduction}

As is well known, smooth SU(N) gauge fields in a 4-dimensional compact
differentiable manifold M have associated an integer topological
charge
\be
Q = - \frac{1}{32 \pi^2} \int_M d^4x \epsilon_{\mu\nu\rho\sigma} 
\tr\left[F_{\mu\nu}F_{\rho\sigma}\right]
\ee
where $F_{\mu\nu}$ is the gauge potential.

This is not merely a mathematical curiosity; it plays a fundamental
role in the understanding of the $U_A(1)$ problem through the anomaly
\cite{Anomaly1,Anomaly2,THooft}, and the Witten-Veneziano formula
\cite{Witten,Veneziano}. It is also crucial for the investigation of
the $\theta$ vacuum in QCD and the strong CP problem \cite{Axion1},
and therefore with the current experimental searches for axions
\cite{Axion2,Axion3}.

The issue of obtaining a theoretically sound and practical definition
for the topological charge of lattice gauge fields is an old
one. Several definitions exist, each with its own advantages and
disadvantages. Some such definitions are purely gluonic, essentially
transcribing the continuum definition to the lattice, whereas others
take advantage of the index theorem and compute the charge as the
index of a conveniently chosen fermionic operator.

In \cite{Adams1}, Adams introduced a new definition of the index of a
staggered Dirac operator, based on the spectral flow of a related
hermitian operator. Some numerical results were obtained there for
synthetic configurations in the $2D$ U(1) model.

The purpose of this paper is to study systematically Adams' definition
in 4D (quenched) QCD (preliminary results were presented in
\cite{latt11}).

\section{Definition of the topological charge}

The basic observation of Adams in \cite{Adams1} is that when
considering the spectral flow definition of the topological charge in
the continuum, there is some freedom in the choice of the relevant
hermitian operator. By a suitable choice, we can construct an operator
which, when implemented in the lattice with a staggered dirac operator
has all the required properties.

In the continuum, for a given gauge field, one usually considers the
spectral flow of the hermitian operator
\be
H(m) = \gamma_5 \left(D - m\right)
\label{H1}
\ee
as a function of $m$. Because of the key property that
\be
H(m)^2 = D^\dagger D + m^2, 
\label{keyproperty}
\ee
if we trace the flow of eigenvalues $\left\{\lambda(m)\right\}$ of
$H$, the ones corresponding to the zero modes of $D$, and only those,
will change sign at the origin $m = 0$, each with a slope $\pm 1$
which depends on the chirality of the corresponding mode. This gives
us the index of $D$, and through the index theorem, the topological
charge $Q$ of the corresponding gauge configuration.

On the lattice, we can substitute in (\ref{H1}) $D$ by the discretized
Wilson Dirac operator
\be
H_W(m) = \gamma_5 \left(D_W - m\right)
\ee
Now the index can be obtained similarly to the continuum, by counting
the number of eigenvalues of $H(m)$ that change sign close to the
origin $m = 0$, taking into account the slope of such crossings. If we
try to do the same with the lattice staggered Dirac operator $D_{st}$,
we realize that the procedure does not work anymore, as the
corresponding $H$ fails to be hermitian.

The key innovation in Adams \cite{Adams1} is the realization that one
can use a different $H$ in the continuum to accomplish the same task,
namely
\be
H(m) = iD - m \gamma_5.
\label{H2}
\ee
This operator is hermitian and verifies (\ref{keyproperty}), and
therefore its spectral flow also gives the index of $D$. But now we
can substitute in (\ref{H2}) $D$ by the lattice staggered
discretization,
\be
H_{st}(m) = iD_{st} - m \Gamma_5,
\label{Hst}
\ee
where $D_{st}$ is the massless staggered Dirac operator and $\Gamma_5$
is the taste-singlet staggered $\gamma_5$ \cite{Golterman}. This
operator is hermitian, and we can study its spectral flow,
$\lambda(m)$. The would-be zero modes of $D_{st}$ are identified with
the eigenmodes for which the corresponding eigenvalue flow
$\lambda(m)$ crosses zero at low values of $m$, and the chirality of
any such mode equals (with our conventions) the sign of the slope of
the crossing \cite{Adams1}. 

Any staggered discretization of the Dirac operator can be used, in
principle, to implement \ref{Hst}. We have chosen to work with the
unimproved, 1-link staggered Dirac operator \cite{KS}, and with the
highly improved HISQ discretization \cite{hisq}. In each case we have
calculated, using standard numerical algorithms, the smallest (in
absolute value) 20 eigenvalues of $H_{st}(m)$ for enough values of $m$
to allow us to determine unambiguously the cuts with the x axis. To
compare with previous work, we have also calculated the low-lying
modes of the HISQ Dirac operator at $m = 0$, and identify the would-be
zero modes with the high taste-singlet chirality ones
\cite{top1,top2}.

\section{2D U(1)}

We started by studying the behavior of the spectral flow of
$H_{st}(m)$ corresponding to the 1-link staggered Dirac operator in 2D
U(1) lattice gauge theory; our purpose was twofold: testing our
numerical framework in a less demanding setting, and working in a
theory with a simple geometrical definition of the topological charge
(even if not immune from problems arising from dislocations
\cite{Teper}). To this end we can either construct gauge fields
configurations with an assigned topological charge \cite{instanton},
or generate realistic field configurations from a canonical ensemble,
selecting those with the charge $Q$ we are interested in.  We used the
first possibility in the earlier stage of our tests and then we moved
to the other option: the results shown in what follows are from this
second phase\footnote{See \cite{Durr1} for related work in the
  unquenched case.}.

We have generated a large ensemble of quenched configurations at
different values of $\beta$, varying from 4 to 9, in order to cover
the scaling region for the lattice size we consider (L=60). Then we
select among this set of configurations subsets of fixed charge $Q$,
for which we computed the spectral flow. 

The spectrum of (\ref{Hst}) has the exact symmetry $\lambda(m)
\leftrightarrow - \lambda(-m)$, therefore we only need to calculate
the flow for, say, $m > 0$. An equal number of crossings, with
identical slopes, will be present for $m < 0$.

\begin{figure}[h]
\includegraphics[scale=1.3,angle=0]{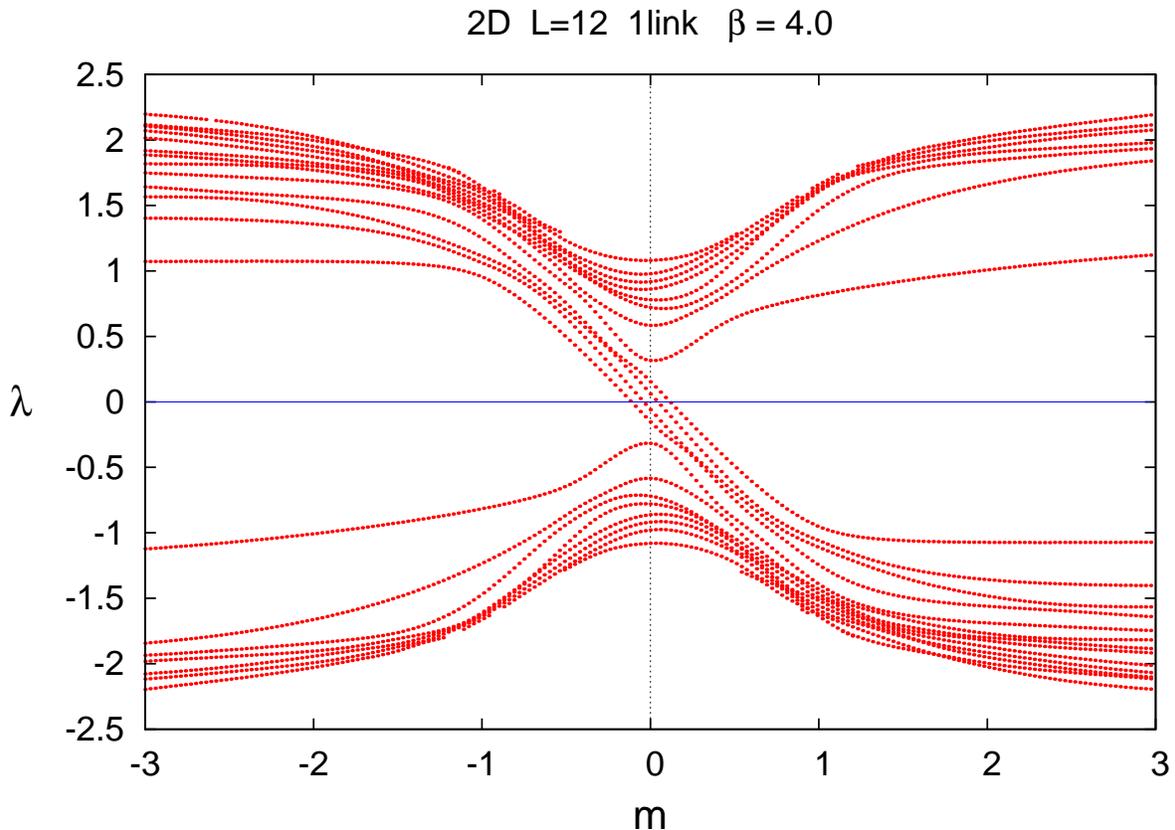} 
\caption{Spectral flow of the 2D 1link Dirac operator
  for a gauge configuration with $Q = -2$. 
\label{fig_2d_1}}
\end{figure}

In Fig. \ref{fig_2d_1} we show the spectral flow for a gauge
configuration with a charge $Q = -2$, corresponding to a $12^2$
lattice and a coupling constant $\beta = 4.0$. We plot the lowest (in
absolute value) 20 eigenvalues of $H(m)$ in the range $(-3, 3)$. We
can appreciate two crossings with negative slopes for $m < 0$, and the
symmetric ones for $m > 0$\footnote{Staggered fermions in $D$
  dimensions have a taste degeneracy of $2^{D/2}$, and therefore the
  number of crossings in the continuum gets multiplied by that factor
  in the lattice.}.

\begin{figure}[h]
\includegraphics[scale=1.3,angle=0]{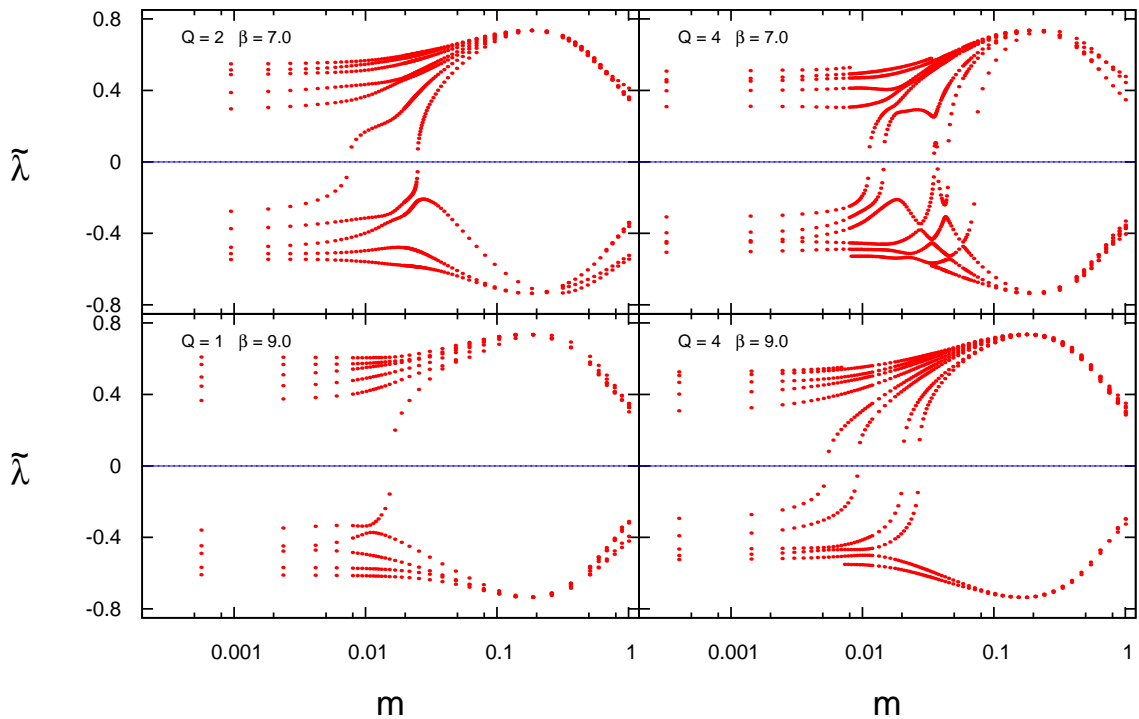} 
\caption{Comparison of the spectral flow of the 2D 1link Dirac
  operator for several gauge configurations.
\label{fig_2d_2}}
\end{figure}

In Fig. \ref{fig_2d_2} we show the spectral flows corresponding to
several gauge configurations in a larger volume, $60^2$, and at
several values of the coupling. From now on we only plot the flow for
$m > 0$. For clarity, in most of the figures we plot the variable
\be 
\tilde{\lambda} = \sgn(\lambda) \sqrt{|\lambda|} \log(|\lambda|) 
\ee
versus $\log(m)$. As we can see in every case we have the expected
number of crossings.

\section{4D Quenched QCD}

For our numerical calculations in 4D we have used three ensembles of
tree-level Symanzik and tadpole improved quenched QCD at three values
of the coupling constant $\beta$ (5.0, 4.8 and 4.6), corresponding
respectively to lattice spacings of approximately 0.077, 0.093 and
0.125 fm \cite{top2}, with lattice volumes of $20^4$, $16^4$ and
$12^4$ respectively . These ensembles are thus approximately matched
in physical volume, $\approx 1.5 \textrm{fm}^4$.

\begin{figure}[h]
\includegraphics[scale=1.3,angle=0]{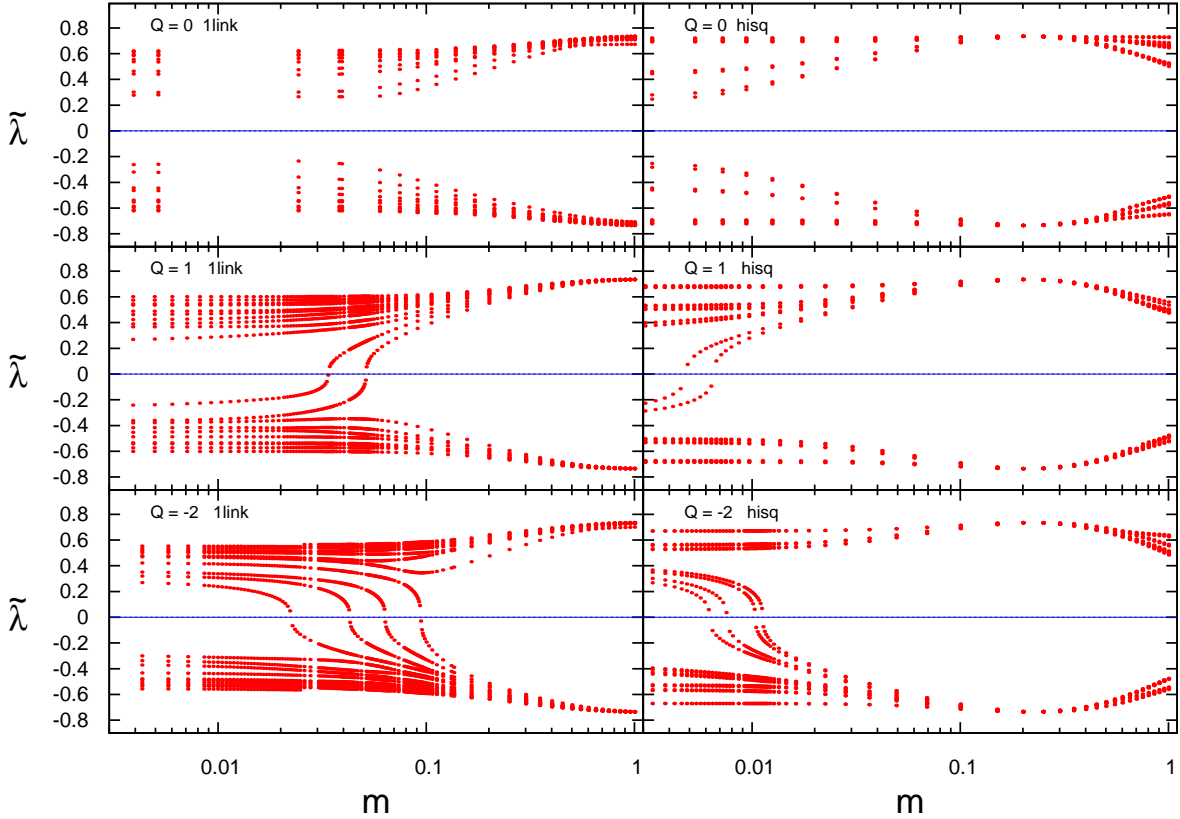} 
\caption{Spectral flow for several configurations at $\beta = 4.6$
  (coarse ensemble) and for various values of the topological
  charge. The left and right panels correspond to the spectral flow
  calculated on the same configuration but with a different operator.
\label{fig_4d_beta46}}
\end{figure}

\begin{figure}[h]
\includegraphics[scale=1.3,angle=0]{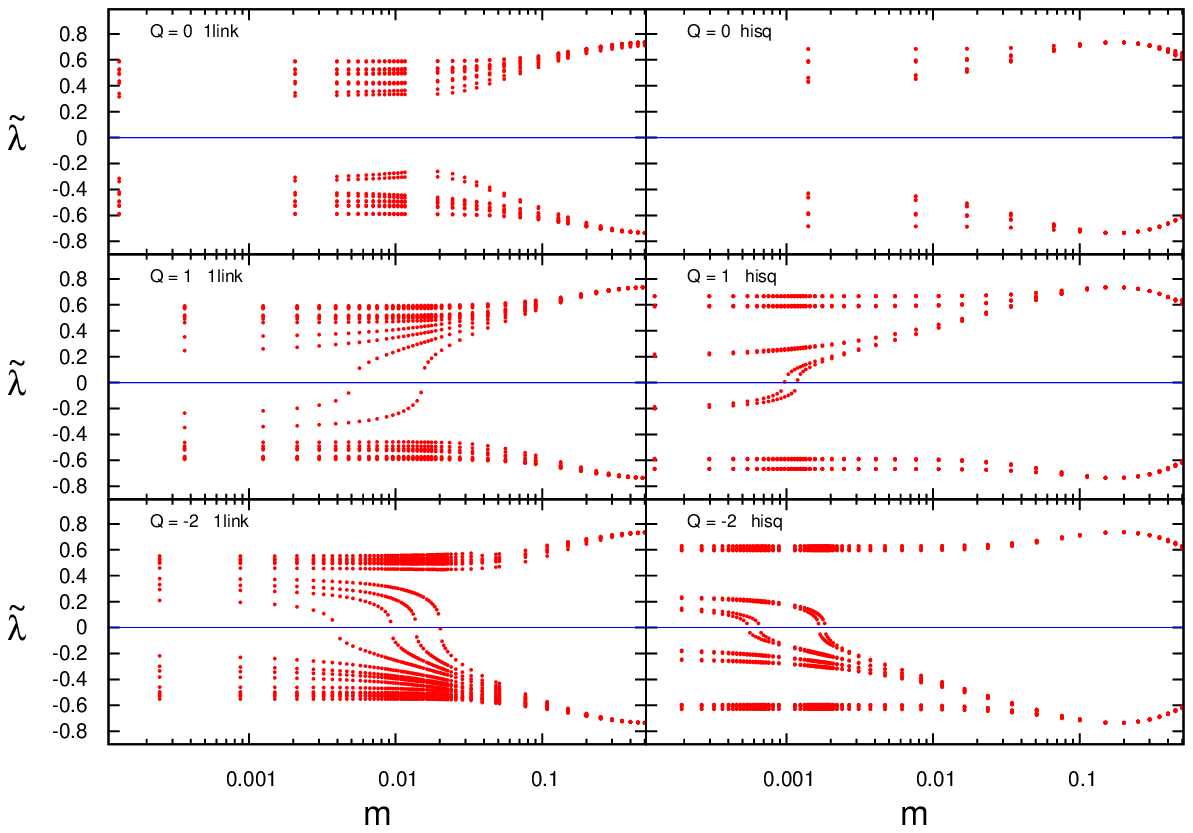} 
\caption{Spectral flow for several configurations at $\beta = 5.0$
  (fine ensemble) and for various values of the topological
  charge. The left and right panels correspond to the spectral flow
  calculated on the same configuration but with a different operator.
\label{fig_4d_beta50}}
\end{figure}

\begin{figure*}[h]
\includegraphics[scale=.45,angle=0]{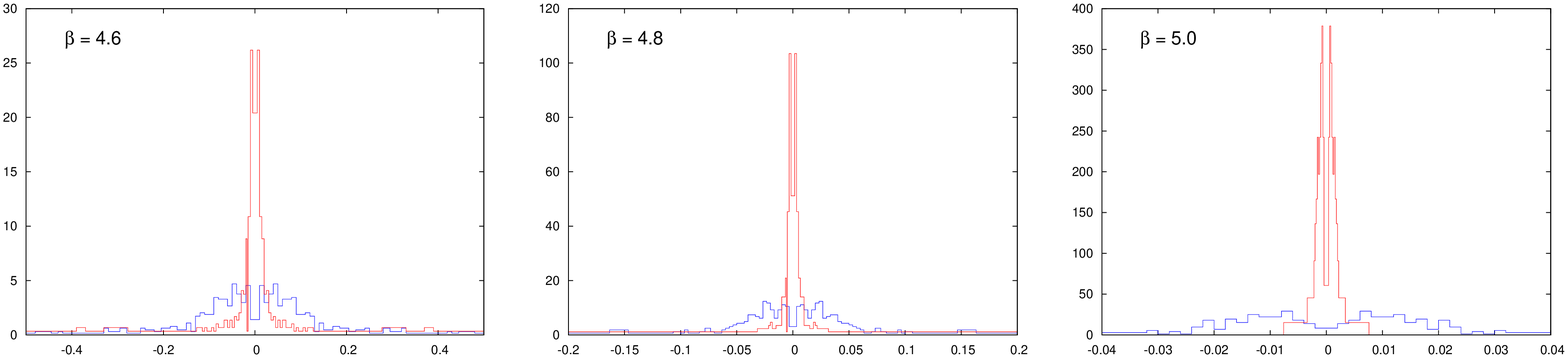} 
\caption{Normalized histograms of the distribution of cuts of the
  spectral flow with the zero x axis for the three ensembles. In each
  individual figure we have plotted both the 1link result (broad
  distribution) and the HISQ result (narrow distribution) on the same
  scale for comparison. Scales are different for the different
  figures.
\label{fig_histograms}}
\end{figure*}

In Figs. \ref{fig_4d_beta46} and \ref{fig_4d_beta50} we show the
spectral flow for several representative configurations from the
coarsest and finest ensembles, and in each case for $H_{st}$ built
from both the 1-link and the HISQ Dirac operators. As before, we
calculate the first 20 eigenvalues in absolute value. The topological
charges were also calculated by counting the number of eigenvectors of
the HISQ Dirac operator with high chirality \cite{top2}, and in each
case there is agreement between the two definitions and for the
spectral flow corresponding to both operators.
We can appreciate in the figures that the cuts of the spectral flow
with the x axis are closer to the origin $m = 0$ for the HISQ than for
the 1-link operators, and also get closer as we go to smaller lattice
spacings. This is consistent with the expectation that in the
continuum limit the cuts should move to the origin, and that the
improved Dirac operator is closer to the continuum than its unimproved
counterpart. In order to make a more quantitative statement, we have
computed a histogram (normalized to area one) of the cuts for the
three different ensembles and both operators, which is shown in
Fig. \ref{fig_histograms}.
We can see clearly the large differences between both operators, and
how the distribution of cuts moves towards zero as we decrease the
lattice spacing.

\begin{figure}[h]
\includegraphics[scale=1.2,angle=0]{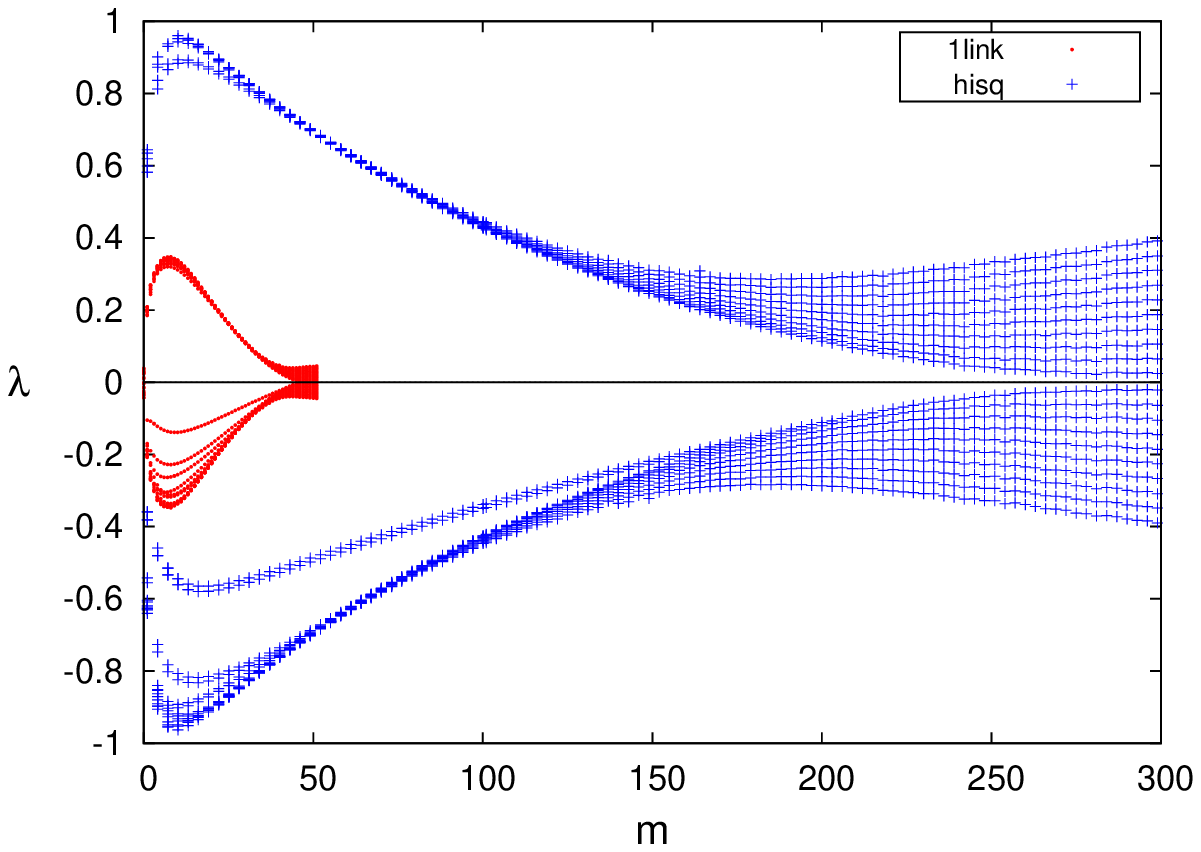} 
\caption{Spectral flow for very large values of m for a typical
  configuration in the fine ensemble, both for the 1link and the HISQ
  staggered operators.
\label{fig_4d_long}}
\end{figure}

In order for the identification of the topological charge through the
spectral flow to be unambiguous, it is necessary that there is a clear
separation between cuts close and far away from the origin. To test
whether this is the case we have calculated a few flows up to very
large values of the parameter $m$. We show in Fig. \ref{fig_4d_long}
a representative result corresponding to a fine configuration, for
both the unimproved and the HISQ cases. We can see that there is a
very clear separation between cuts close to the origin and other
possible cuts, which are very far away from the origin for both
operators.

\begin{figure}[h]
\includegraphics[scale=1.3,angle=0]{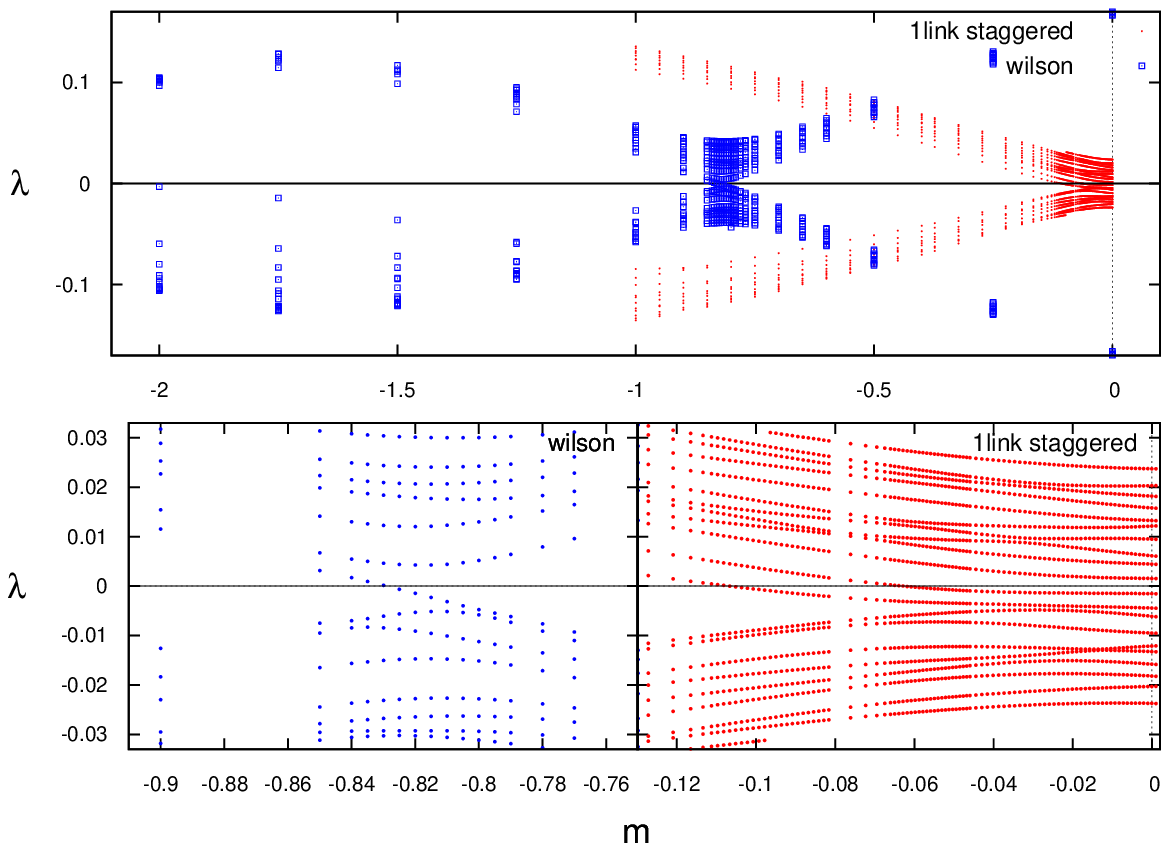} 
\caption {Comparison of the Wilson and the staggered flow for the same
  configuration of charge $Q = -1$ in the coarse ensemble. The lower
  figures are a detailed view of the crossings with the x axis.
\label{fig_4d_wilson_beta46}}
\end{figure}

\begin{figure}[h]
\includegraphics[scale=1.3,angle=0]{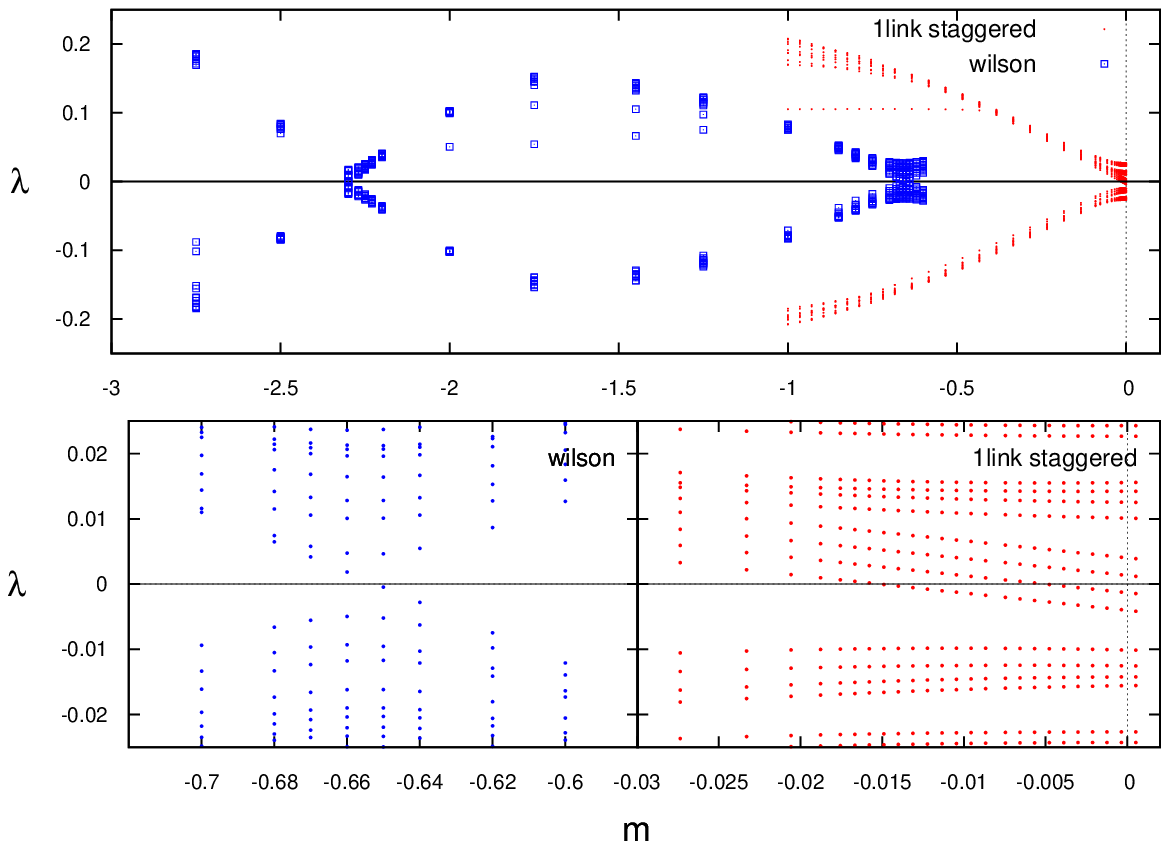} 
\caption {Comparison of the Wilson and the staggered flow for the same
  configuration of charge $Q = -1$ in the fine ensemble. The lower
  figures are a detailed view of the crossings with the x axis.
\label{fig_4d_wilson_beta50}}
\end{figure}

We have also calculated the spectral flow corresponding to the Wilson
Dirac operator for a few configurations, in order to compare the
results with the staggered case. We have chosen to do the comparison
with the flow corresponding to the 1-link operator, which is
unimproved, and therefore in a sense a closer relative to the Wilson
one. In Figs. \ref{fig_4d_wilson_beta46} and
\ref{fig_4d_wilson_beta50} we show both the Wilson and the 1-link
staggered flow for two configurations corresponding to the coarsest
and finest ensembles, both with $Q = -1$. As we can see we get
consistent results in both cases. On the other hand, the computer time
needed for the calculation is considerably less in the staggered case,
possibly due to the better conditioning of the staggered Dirac
operator.

\section{Conclusions and Outlook}

We have presented clear numerical evidence that Adams' definition of
the topological charge using the staggered Dirac operator works as
expected also for realistic (quenched) $SU(3)$ gauge fields. The
crossings near and far away from the origin are very well separated,
and therefore the topological charge of a configuration is
unambiguously defined, even in cases which would be ambiguous using
other definitions. For most configurations we have seen that the
charge as measured by the number of high taste-singlet chirality modes
and by the spectral flow agree. We have also seen the expected
differences between the position of the cuts between the 1link and the
HISQ operators, as well as the clear move towards zero of the cuts as
we decrease the lattice spacing.

For a future work, it would be interesting to repeat this study in
full QCD ensembles, including the effect of sea quarks.

Inspired by this construction, it is possible to define an overlap
operator starting with a staggered kernel, instead of the usual Wilson
one \cite{Adams2}, producing a chiral operator representing two tastes
of fermions. A similar construction can be carried out to further
reduce the degeneracy and produce a one-flavor overlap operator
\cite{Hoelbling}. The question is whether this construction has all
the required properties, and is further numerically advantageous as
compared with the usual overlap construction. Results are presented in
\cite{deForcrand1,deForcrand2,Durr2,Adams3}.

\begin{acknowledgments}

We thank Alistair Hart for generating the configurations.  This work
was funded by an INFN-MICINN collaboration (under grant
AIC-D-2011-0663), MICINN (under grants FPA2009-09638, FPA2008-10732
and FPA2012-35453, cofinanced by the EU through FEDER funds),
DGIID-DGA (grant 2007-E24/2), and by the EU under ITN-STRONGnet
(PITN-GA-2009-238353). E. Follana was supported on the MICINN Ram\'on
y Cajal program. and A. Vaquero was supported by MICINN through the
FPU program. E. Follana acknowledges financial support from the
Laboratori Nazionali del Gran Sasso during several research visits
where part of this work was carried out.

\end{acknowledgments}

\end{document}